\documentclass[a4paper,twoside]{article}

\usepackage{epsfig}
\usepackage{subcaption}
\usepackage{calc}
\usepackage{amssymb}
\usepackage{amstext}
\usepackage{amsmath}
\usepackage{amsthm}
\usepackage{multicol}
\usepackage{pslatex}
\usepackage{natbib} 
\usepackage{algorithm2e}
\usepackage[bottom]{footmisc}

\usepackage[hyphens]{url}
\usepackage{multirow}
\usepackage{csquotes}

\usepackage{SCITEPRESS}     

\begin{document}

\title{Developers Insight On Manifest v3 Privacy and Security 
Webextensions}


\author{\authorname{Libor Polčák\sup{1}\orcidAuthor{0000-0001-9177-3073}, Giorgio Maone\sup{2}, Michael McMahon\sup{3}, and Martin Bednář\sup{1}}
\affiliation{\sup{1}Brno University of Technology, Faculty of Information
Technology, Božetěchova 2, 612 66 Brno, Czech Republic}
\affiliation{\sup{2}Hackademix, via Mario Rapisardi 53, 90144 Palermo, Italy}
\affiliation{\sup{3}Free Software Foundation, 31 Milk Street, 960789 Boston, MA 02196 USA}
\email{\{polcak, ibednar\}@fit.vut.cz, giorgio@maone.net, michael@fsf.org}
}

\hyphenation{Chro-mi-um web-ex-ten-sions web-ex-ten-sion par-ti-ci-pants
res-pon-dents}


\abstract{
  Webextensions can improve web browser privacy, security, and user
  experience.
  The APIs offered by the browser to webextensions affect possible
  functionality. Currently,
  Chrome transitions to a modified set of APIs called Manifest v3. This paper
  studies the challenges and opportunities
  of Manifest v3 with an in-depth structured qualitative research.
  Even though some projects observed positive effects, a majority
  expresses concerns over limited benefits to users, removal of crucial APIs, or
  the need to find workarounds. Our findings indicate that the transition
  affects different types of webextensions differently; some can migrate without
  losing functionality, while other projects remove functionality or decline to
  update. The respondents identified several critical
  missing APIs, including reliable APIs to inject content scripts, APIs for
  storing confidential content, and others.
}

\onecolumn \maketitle \normalsize \setcounter{footnote}{0} \vfill

\section{\uppercase{Introduction}}

Webextensions are a collection of Javascript code run by a web browser.
%
As web and browsers evolve, requirements change, and occasionally,
extension developers need to adapt their code~\citep{why_xul_removed}.

At the end of 2018, Google proposed a breaking change in the webextension API
called Manifest v3 (Mv3) that changes several crucial mechanisms of
webextensions~\citep{google_mv3_2018}.
The goal of Mv3 is to enable the easy creation of secure, performant, and
privacy-respecting extensions, while writing an insecure, non-performant, or
privacy-leaking extension should be difficult~\citep{google_mv3_2018}.
According to Google,
Mv3 webextensions should have the same capabilities as Manifest v2 (Mv2)
webextensions\footnote{\url{https://groups.google.com/a/chromium.org/g/chromium-extensions/c/qFNF3KqNd2E/m/uZlWrml1BQAJ}}.
However,
the proposal quickly received
criticism~\citep{forbes_mv3_2019_timeline,chrome_stats_mv3}.


We started this research to shed light on the dispute between Google and the
critiques.
Specifically, we aimed at answering research questions: (RQ1)~How are webextension
projects affected by the migration from Mv2 to Mv3? (RQ2)~Are developers reluctant
or hesitant to migrate? (RQ3) Is the migration smooth? (RQ4)~Does Mv3 make webextensions better?
To find answers to these questions, we conducted the structured
qualitative study complemented with the longitudinal study based on
\url{chrome-stats.com}.

Due to the diversity of webextensions and the distinct impact of Mv3 on
various APIs offered to webextensions, we observe miscellaneous consequences. Some
webextensions could have been updated quickly. Other projects report a migration period
of up to one year. Some projects decided not to migrate as Mv3 lacks
essential APIs or due to frustration from frequent forced changes or changing
deadlines. Other projects remove functionality due to missing APIs.
Most participants do not consider Mv3 to make writing safe and
privacy-friendly
webextensions easier as participants need to spend time to find workarounds
around missing or buggy APIs.

We disclosed all research details to our participants.
We stressed that participation in the research is
voluntary. Consequently, approval from a research ethics board is not
required.

This paper is organized as follows.
Section~\ref{sec:differences_23} describes differences between Mv2 and Mv3.
 Section~\ref{sec:related} reviews related work. Section~\ref{sec:study}
 introduces our methodology and the participants. Section~\ref{sec:results}
 reports the results and discusses their impact. The paper is concluded by
 section~\ref{sec:conclusion}.

\section{\uppercase{Necessary Background}}
\label{sec:neceassary}

This section explains what webextensions are, their internals, what
Manifest is, and how its version impacts webextensions.

\subsection*{The Nature of Webextensions}

Webextensions are a collection of Javascript code run by the browser. A user
installs webextensions manually, typically using dedicated pages
operated by browser vendors\footnote{For example,
\url{https://addons.mozilla.org/} for Firefox and Chrome Web Store
\url{https://chromewebstore.google.com/} for Chrome.}.
Originally, webextensions appeared in browsers derived from Chromium. Firefox
later adopted similar APIs~\citep{why_xul_removed}.
After a user installs a
webextension to a browser, the browser starts the webextension with every
browser launch.
Webextensions can perform tasks that are not possible for scripts of visited
pages, like extending the browser UI, modifying the content of all or
selected visited pages, and other actions, some of which can significantly
change the appearance and content of visited web pages. Except for benign usage, the
APIs were observed to be used by malicious code~\citep{spy_little_eye,DoubleX}.

Even though browser vendor tend
to provide APIs compatible to other browsers to ease the development of
webextensions for multiple browsers\footnote{\url{https://extensionworkshop.com/documentation/develop/porting-a-google-chrome-extension/},
\url{https://developer.apple.com/documentation/safariservices/safari_web_extensions}},
differences exists\footnote{For example, DNS API is available only in Firefox. \url{https://developer.mozilla.org/en-US/docs/Mozilla/Add-ons/WebExtensions/API/dns}
}.
Consequently, the W3C WebExtensions Community Group
works on standardizing the APIs. One of the principles is compatibility, which
aims to specify APIs that should be available in all browsers. However, no
standard exists yet.

The variety of available APIs for webextensions creates opportunities to develop
diverse tools.
Some focus on enhancing the possibilities of
browsers or specific pages, like extending the download options, providing
advanced options to browser UI like enriching the possibilities of working with
opened tabs.
This paper focuses on webextensions improving the privacy and security
of web users.

The privacy webextensions usually deal with web
tracking~\citep{web_never_forgets,internet_jones,block_me_if_you_can}, typically
blocking some web requests (network communication initialized by the browser due
to the content of the visited web page) or parts of the visited web page (for
example, cookie banner removals remove forms trying to obtain consent with
tracking or page content sanitizers remove third party content displayed on
pages that is not necessary for the page like social network plug-ins).

Security webextensions improve the security mechanisms offered by browsers. For
example, password managers allow users to set strong and unique passwords,
webextensions can help encrypt messages (end-to-end encryption) or guard
network boundaries so that a browser cannot be misused as a
proxy~\citep{forcepoint,jshelter}.

\subsection*{Webextension Internals}

A webextension typically consists of several components.
For example, a
webextension might (1)~need to perform some computation in the background for each visited
page or share data for multiple visited pages, (2)~deploy scripts to visited
pages, or (3)~enrich the browser UI with additional icons or menu entries
so the user can interact with the webextension.
Often the components need to communicate with each other.

As background scripts, content scripts, and scripts in popup windows each run
in a different context and cannot directly share information, each communication
needs specific means. Webextensions offer APIs to communicate with
messages exchanged between the components of the webextension.

\section{\uppercase{Mv2 and Mv3 Differences}}
\label{sec:differences_23}

Webextensions 
contain a file \verb|manifest.json| with
metadata about the webextension, such as its name or required permissions.

Browser vendors version the Manifest to adapt to the evolving needs.
Consequently, the browser can run a webextension
differently based on the Manifest version.
Mv3 builds on top of the previously used Mv2 and maintains most
of the APIs available to webextensions. However, some APIs and mechanisms
available to webextensions were redesigned.

Webextensions often need to maintain data. For example,
a user can configure and fine-tune the functionality of many webextensions, or
webextensions need to store computed information about visited
pages or the browser as a whole\footnote{For example, the number of
blocked trackers on the currently visited page.}. Both Mv2 and Mv3 allow
storing state in a database in the browser. However, the space is
limited\footnote{\url{https://developer.mozilla.org/en-US/docs/Mozilla/Add-ons/WebExtensions/API/storage/session}}.
Consequently, Mv2 webextensions stored state in \emph{background pages}, which
offered a private Javascript run time environment spanning whole browser session
in which the webextension can perform any computation or store any data~\citep{google_mv3_2018}.
Mv3 replaced the component with \emph{background service workers} that can stop
running and respawn at any time. The motivation is to
lower memory requirements as the memory is freed when the worker stops running.
However, the shift from a persistent to the short-lived environment may affect
the capabilities of
webextensions\footnote{\url{https://groups.google.com/a/chromium.org/g/chromium-extensions/c/Qpr-Wf8BcsI/m/hC0RNrvqCwAJ}}.

Webextensions can observe web requests made by the browser. \emph{Blocking web
requests}
allow webextensions to read and modify the content of the requests and
the responses. Various blockers have used this functionality to
block some requests completely or to sanitize the replies.
Non-blocking web requests allow only reading the messages. Mv3 removed
blocking web requests and replaced them with declarative net request
API\footnote{\url{https://developer.mozilla.org/en-US/docs/Mozilla/Add-ons/WebExtensions/API/declarativeNetRequest}}.
\emph{Declarative net requests} allow webextensions to specify rules in advance that
(1)~define web requests that should be modified (for example, by selecting
specific domains or by applying regexes on URLs), and (2) define actions on the
selected requests like blocking, redirecting, or modifying the message.
Nevertheless, blocking web requests API is more powerful as the webextension could
have used any data in the observed messages and extension state to decide the
actions applied to the message. For declarative web requests, the computation
needs to happen in advance, lacking data current to the exchanged message and
webextension internal state~\citep{eff_google_plans}. In addition, the number of declarative rules is
limited.

\section{\uppercase{Related Work}}
\label{sec:related}

Whereas some online resources already provide feedback from developers of
webextensions~\citep{eff_mv3_deceitful,chrome_stats_mv3}, this paper is the first to
systematically study the attitude of the developers and the obstacles associated
with the migration like the quality of documentation, ease of development, and
the maturity of the APIs as well as the longitudinal trends in the migration.

\citet{manifest_v3_unveiled} study the effect of
Mv3; however, they do not
learn first-hand experience from the developers and do not focus on privacy- and
security-related webextensions. They see 87.8\,\% removal of APIs related to
malicious behavior in Mv3. However, after adaptation, 56\,\% of the
examined malicious webextensions retain their malicious capabilities within the
Mv3 framework. Whereas they found that Google refines the APIs to
address developers' concerns, we show that these refinements are insufficient and a
significant portion of webextensions does not migrate. Our research
participants often explain that missing APIs are the reason not to migrate.

\citet{webextension_developers_advocacy} also performed a
qualitative analysis of developers' attitudes to the development of privacy
webextensions. However, they focused on the motivation of the developers to
create this kind of software.

We are not the first to use \url{chrome-stats.com} data for a long-term
analysis. 
\citet{chrome_web_store_longitudinal} employed data
from \url{chrome-stats.com} to 
provide a holistic view of the webextensions
listed in Chrome Web Store and their lifecycle.
The work of Sheryl Hsu et al. is more generic than this paper.
Nevertheless, they do not deeply study and explain the issues of the Mv3 transition.
Moreover, their paper was written before Google released the first Chrome
that turns off Mv2 webextensions by default. We are the first to
show the effect of the deadline.

We observe uncertainties about the effectiveness of the APIs (see, for example,
section~\ref{sec:content_scripts}). While previous studies~\citep{watching_them_watching_me}
unveiled that some users feel
protected by webextensions, we show that developers of webextensions are not certain about the correct
behavior of their webextensions.

Without a doubt, webextensions are misused for malicious
purposes~\citep{mystique,DoubleX}
\citep{spy_little_eye,DoubleX,arstechnica_malware_extensions_after_purchase,hulk,hardening_security_extensions,mystique}.
For example, 
\citet{mystique} discovered thousands of
webextensions potentially leaking privacy-sensitive information, some of which
have over 60 million users. 
\citet{DoubleX} detected suspicious data
flows between external actors and sensitive APIs in webextensions and
demonstrated exploitability for 184 extensions.

These security problems led to a debate among browser vendors
about removing some capabilities for extensions, with
the most dangerous capability being the ability to inspect and modify or prevent
any request the browser makes to any website~\citep{mv3_privacy_performance}.
Removing the functionality would clearly prevent malicious extensions from
misusing it, and, hence, protect users’ privacy~\citep{mv3_privacy_performance}.
The removal of blocking web request API goes in this direction.
However, previous work has pointed out that removing blocking web request APIs does
not prevent malicious webextensions from observing all browser traffic~\citep{eff_google_plans}.

\citet{mv3_privacy_performance} benchmarked eight popular privacy-focused browser
webextensions. 
They measured that privacy-focused extensions not only
improve users' privacy but can also increase users'
browsing experience.

Previous work also showed that the declarative net request API is
inferior to the blocking web request API as webextensions cannot make decisions
based on contextual data~\citep{eff_google_plans}.

Some webextensions unintentionally and unnecessarily hinder security. For
example, 
\citet{helping_hindering} detected extensions that
remove security headers 
or 
affect pages on
top of their declared policy. Such modifications continue to be applicable in
Mv3~\citep{helping_hindering}.


\section{\uppercase{The Study}}
\label{sec:study}

To address the research questions, we performed a structured qualitative
research based on questionnaire with open questions sent to webextension developers, see Appendix~\ref{appx:questionnaire},
We also observed Chrome Web Store and tracked published versions. For
missing data, we employed
Chrome-Stats, 
an independent project that
tracks published webextensions in Chrome Web Store.

\subsection{Methodology}
\label{sec:methodology}

Our research goal needs to focus on the first-hand experience to get insight
into the mindset of developers of distinct web extensions. As we did not know
what might be all properties that influence the migration as well as the
obstacles to migration and the qualities that Mv3 brings to the webextension
ecosystem, we opted for qualitative research with open questions to allow the
respondents to express their views freely
(see
Appendix~\ref{appx:questionnaire} for the exact wording of the questions).
To save time for developers and give them enough time to contribute, we
prepared a structured questionnaire\footnote{\url{https://pastebin.com/nTFykFNi}} and asked
participants to reply in one month but allowed them to extend the deadline.
Several
projects asked for a deadline extension.

One of the authors of the paper keyworded all replies to a
vocabulary based on the content of the open answers. Another
author checked the keyworded answers.
During the process, when we had access
to an independent source, we also checked the validity of the
answers (for example, we compared the status of migration to the
published version in Chrome Web Store).
We ignored some answers, for example,
when respondents speculated or did not provide a personal experience for questions
where we sought personal experience.
We provide aggregated
results in Section~\ref{sec:results} accompanied by the most interesting
insights from the questionnaire.

\subsection{Search For Participants}
\label{sec:searching_participants}

We selected privacy- and security-related projects based on
publicly listed webextensions for
Firefox\footnote{\url{https://addons.mozilla.org}} and
Chrome\footnote{\url{https://chrome.google.com/webstore/}}.
We asked 3,041
projects and 10 additional WebExtensions W3C working group participants
questions aiming to answer our research questions in November 2023 and received
the latest reply at the beginning of February 2024.

For Firefox, we studied webextensions listed under the
\emph{Privacy and Security} category\footnote{\url{https://addons.mozilla.org}}.
Hence, we contacted only
developers that self-report that their webextensions deal with privacy or
security. As the listings do not provide contact details to reach
developers, we collected e-mail addresses for support. We gathered e-mail
addresses for 2,159 webextensions listed at \url{addons.mozilla.org}. 1,547
webextensions do not contain a support link in the listing, and we omitted these
webextensions from our research.

At the time we started our research, Chrome Web
Store\footnote{\url{https://chrome.google.com/webstore/}} did not offer
a category for privacy or security webextensions. However,
a visitor to the store can search for keywords. We searched for \emph{privacy}
and \emph{security}, and collected webextensions that the store offered. We
expanded the list as long as the names of the webextensions were
connected to privacy or security. Although we have not curated the list (so it also
contains webextensions that do not deal with privacy or security but were for
some reason offered to us),
we validated the purpose of the webextensions by one of the questions (see Table~\ref{tab:3b}).
Only two participants develop webextensions not related to privacy and security.
We collected 1,110 webextensions 
from Chrome Web Store and found 1,058 e-mail addresses.

Additionally, we asked regular  WebExtensions W3C
Working Group participants as the group comprises of active webextension
developers and
other stakeholders. 
They have first-hand experience in
the standardization process and should be the most familiar with the issues
and opportunities raised by Mv3.

In total, we found 2,702 unique e-mail addresses in \url{addons.mozilla.org} and
Chrome Web Store. 12 of these addresses belonged to the projects of participants of
WebExtensions W3C Working Group, and we excluded these addresses. Hence, we
collected 2,690 e-mail addresses from \url{addons.mozilla.org} and the Chrome Web Store
we used for the interviews.

257 e-mails of  2,690  that we sent to e-mail addresses listed as support in
\url{addons.mozilla.org} or Chrome Store were rejected by the receiving e-mail
servers. Typically, the text of the message indicates that the e-mail address
does not exist. However, in some of the cases, it is not clear if the reason was an
e-mail address that does not exist or if the e-mail was rejected because it was
classified as spam.

We received 33 replies that contained meaningful answers. One of the
participants provided answers for six projects in one answer without sufficient
data to distinguish between the projects. One participant reported two very similar
project names in one answer. We counted each of these two participants just once.
One participant answered for two projects and another for three projects. As the
answers clearly distinguish between the projects, we counted each answer as one
for questions dealing with developers' opinions and counted each project
separately for questions dealing with the projects.

We gave each reply a unique participant project ID (PPID) of R\emph{XX} where
\emph{XX} are two digits. We refer to PPID in the following text to distinguish
the projects or their answers.

Table \ref{tab:3a} shows that we received answers for both small webextensions that
have just a few users as well as answers for webextensions with hundreds of
thousands of
users or more\footnote{\label{ftn:users_count}Previous research reports that the
number of users reported by Chrome Web Store is not
precise~\citep{chrome_web_store_longitudinal} as users not using their computer
longer than a week are not counted. Additionally, users with more browser
profiles might be counted multiple time. We explicitly asked the participants to
provide their estimates. However, often, participants did not have more precise
numbers than Chrome Web Store.}. Consequently, we provide insight into both
small and large projects.

\begin{table}[h]
  \caption{Number of users of the participating extensions}
  \label{tab:3a}
  \centering
  \begin{tabular}{lr}
Users & Participating webextensions\\
    \hline
Undisclosed & 5\\
Few & 1\\
Tens & 3\\
Hundreds & 9\\
Thousands & 3\\
Tens thousand & 5\\
Hundreds thousand & 3\\
Millions & 1\\
\end{tabular}
\end{table}

The participants develop various kinds of webextensions (see Table~\ref{tab:3b}).
We decoded the primary purpose of each webextension in all answers except the
participant R25. R25 is both an ad blocker and data obfuscation tool,
and is counted in both categories.

\begin{table}[h]
  \caption{The purpose of participating webextensions}
  \label{tab:3b}
  \centering
  \begin{tabular}{lr}
Purpose & Count\\
    \hline
Cookie banner removal & 2\\
Page content sanitizer & 1\\
Cookie manager & 3\\
Ad blocker & 3\\
Tracker blocker & 1\\
Other blocker & 3\\
Referer modifications & 1\\
Password manager & 2\\
Authentication tool & 1\\
Checksum validator & 1\\
Message encryptor & 1\\
Data obfuscation & 1\\
Proxy manager & 2\\
Network boundaries separation & 1\\
Security leak detector & 1\\
Video meeting selector & 2\\
Not related to privacy or security & 2\\
Undisclosed & 3\\
\end{tabular}
\end{table}

14 webextensions attract any user.
3 webextensions are suitable for
users of a specific site, 2 webextensions appeal to
customers of a specific company. 4~webextensions aim
at power users that usually have certain IT skills and
1 webextension is for testers. R22 is 
both 
a webextension for power users and a webextension for everybody.
7 projects did not disclose the type of a user attracted
by the webextension.

The majority of webextensions support Chromium-based
browsers (at least Chrome reported), and 3 of these also support at least one
Safari-based browser. While 6 projects support only Firefox, we
did not receive any answer of a developer that supports just Chromium-based
browsers. Nevertheless, 1 project
reported that it would remove Firefox support during the migration to Mv3.
Another project
added support for Firefox during the migration
to Mv3.



We ignored replies that were not in English. 
All look like a confirmation e-mail or are too short to
contain answers to the questions.
We also ignored messages
stating that the
project would not provide answers or
that lacked answers to most questions and merely
stated dissatisfaction with Mv3 based on other webextensions.
We ignored one reply of a
webextension developer that plans to retire the webextension as it is obsoleted by
other webextensions. The developer did not have any practical experience with
Mv3
and did not provide any relevant answers.

Often, we received a confirmation e-mail. A few times, we were asked to confirm
the message by clicking a link or sending another e-mail. In such cases, we
followed and performed such actions. However, we did not create accounts when we
were asked.

\section{\uppercase{Results and Discussion}}
\label{sec:results}

This section provides  an analysis of the responses in the context of
our research questions.

\subsection{RQ1: Are Webextensions Affected?}
\label{sec:affected}

Firstly, we were interested in whether our participants are affected by the Manifest
change and their migration plan (RQ1). Table \ref{tab:4a+f} shows
that the majority of participating web
extensions are affected. One project reported both that it is inactive and
unsure how affected; that project is counted only as inactive.

\begin{table}[h]
  \caption{Migration status (at the end of 2023 or the start of 2024)}
  \label{tab:4a+f}
  \centering
  \begin{tabular}{lrrrrrr}
\parbox[b]{3cm}{Affected by Mv3?} & \rotatebox{90}{\parbox{1.8cm}{Not planned}} & \rotatebox{90}{Not started} & \rotatebox{90}{Exploration} & \rotatebox{90}{Paused} & \rotatebox{90}{In progress} & \rotatebox{90}{Finished}\\
    \hline
Yes & 5 & 0 & 2 & 2 & 7 & 6\\
No & 0 & 1 & 0 & 0 & 0 & 1\\
Likely not & 0 & 1 & 0 & 0 & 0 & 0\\
Not sure how & 1 & 1 & 0 & 1 & 0 & 0\\
Uses Mv3 from start & 0 & 0 & 0 & 0 & 0 & 1\\
Project inactive & 0 & 1 & 0 & 0 & 0 & 0\\
\end{tabular}
\end{table}

4 projects do not plan
to migrate the Chrome version of the extension and they want to continue the
extension only for Firefox. One project plans a complete shutdown, and one is unsure if
the author would bother with the migration.
2 paused the migration as
there are missing APIs, whereas one project seems confused about the
scope of changes and frustrated with the impact of the changes on the users.
Whereas 6 projects declare a full transition to Mv3,
2 projects plan to keep the code
for Firefox in Mv2. For example, one project reported compatibility
issues.

The rest of this
subsection considers only the 22 webextensions that
self-reported to be affected as these projects reported meaningful answers to
questions regarding features that were added, lost, or needed to be rewritten due
to the Manifest change.

19 projects did not report any new feature stemming from
Mv3. Only 2 projects
reported that the webextension gains availability on low-resource systems, and a
single project 
reports that it became compatible with Android during the migration.
Additionally,
R07 reports that they reimplemented their webextension to Mv3, and the new
declarativeNetRequests allowed the extension to interact with web requests less
compared to the web request API --- consequently, the extension improved performance.

Table \ref{tab:4d} shows that
almost half of the webextensions do
need a rewrite. Most often, the change in handling background scripts
triggers the need for a rewrite. Only
2 projects declared that a major redesign of
the webextension is needed. One of the respondents reported that he needed to
push for a missing API that browser vendors later added. 

\begin{table}[h]
  \caption{What needs to be rewritten?}
  \label{tab:4d}
  \centering
  \begin{tabular}{lr}
Response & count\\
    \hline
Undisclosed, probably nothing & 13\\
Background scripts & 6\\
Migration to declarative net requests & 3\\
Initialization code & 1\\
Communication to content scripts & 1\\
Communication to pop-up window scripts & 1\\
External library that is not migrated & 1\\
Huge redesign & 2\\
\end{tabular}
\end{table}

One respondent shared that they need persistent background pages and found a
workaround that prevents the browser from stopping the background pages. They
found that the service worker runs forever if they initiate a periodic request
to local storage.

Only 2 projects stated
that they lost functionality due to migration; in both cases, due to dependency on blocking web request API. However,
6 projects decided not to migrate. Some additional
projects did not migrate in time, even though the projects anticipated migration
in their answer.

\subsection{RQ2: Obstacles in Migration}
\label{sec:migration_problems}

RQ2 focuses on the reasons that make the projects hesitant or
even reluctant to migrate.

\subsubsection{Missing APIs}
The most obvious reason complicating or even preventing the migration is
missing APIs.
3 projects lack
APIs simplifying work with non-persistent background workers, and
1~project considers
the background service workers buggy. The lack of blocking web requests
impacts 6 projects, including one
project that self-reported that it was still determining the exact scope of the
impact. One project reports that permissions work differently in Firefox and
Chrome. 2 projects consider
that Mv3 APIs need more real-world experience and are not mature enough.

Some developers
actively highlight missing APIs to browser vendors. For example, R07 was
successful and migrated after addition of a new API.

\subsubsection{Stability and maturity of APIs}

Table \ref{tab:5a} shows that the respondents disagreed on the maturity and
stability of the APIs.
We attempted to correlate the satisfaction
with the APIs' maturity and stability to project purpose or migration status but
found no correlation. Interestingly, similar projects have varying opinions on
this matter.

\begin{table}[h]
  \caption{Satisfaction with the maturity and stability of APIs}
  \label{tab:5a}
  \centering
  \begin{tabular}{lr}
Response & count\\
    \hline
Undisclosed & 5\\
Not studied the APIs & 8\\
No opinion & 3\\
No & 7\\
Yes, but problems with Firefox & 1\\
Yes, but some use cases missing & 1\\
Yes & 4\\
Manifest v3 APIs are more polished & 1\\
\end{tabular}
\end{table}

\subsubsection{Quality of documentation and debuggers}

Most participants who expressed their views on the documentation quality
were positive (13 projects) rather than negative
(3 projects). Yet, the participants expressed that
they lack more examples. Also, participants expressed worries about whether the
documentation is up-to-date due to changes in the APIs and expressed uncertainty
about the stability of the APIs as the documentation does not highlight APIs
that can change. Some projects expressed difficulties in navigation and a lack of
comprehensibility.

R01 gives an example of broken documentation in oversimplifications.
Developers are expected to replace
\verb|setTimeout| calls with Alarms API\footnote{See
\url{https://developer.chrome.com/docs/extensions/develop/migrate/to-service-workers}
for the advice.}. However, R01 found that it is more complex, and the developer
must carefully decide the correct approach.


Most participants did not express their views on debugging tools. However,
5 projects find debugging tools good. Some projects
consider the debugging tools difficult to use, for example, in the context of
debugging imported libraries.

\subsubsection{Browser compatibility}

Another possible reason that might prevent projects from migrating is code
split. Only R35 reported that the project used different code bases for Firefox
and Chrome, and the project plans to merge the code bases during migration to
Mv3. Additionally, R23 finished the migration, and while the project did
not support Firefox before the migration, it added the support during the
migration. 2 projects plan to keep the same code base for
all supported browsers after migration, and one project inserted browser-specific
workarounds to the code base during the migration.

Other projects are going in
the negative direction: 6 projects started to
use different code bases during the migration, additional
6 projects do not plan migration
(meaning that the projects drop support for Chrome), and
1 project
dropped support for Firefox.

R34 warns that the uncertainties in deadlines and support mean the
project might need to migrate twice: first, for Chrome, and later
for Firefox. That means that some decisions and
development time might be invested unnecessarily.

R19 
postponed
migration until it becomes unavoidable. Although the participant
believes the extension should be compatible with Mv3, he is only
interested in migrating once Firefox forces him to do so. R19 is prepared to
lose all Chrome users. This decision makes sense for Firefox users as it allows
the author to focus on other aspects of the extension that directly impact them.

%
We asked if Mv3 improves the compatibility of webextensions
across browser vendors. One participant did not know that Mozilla
works on Mv3 support for Firefox. While counting this project, 14
projects expressed a negative opinion. R08 and R19 highlight that Mv3,
as implemented by Chrome, removes some functionality still available in
other browsers, making the compatibility worse.
Nevertheless, eight participants expressed
positive answers. One of these is R07, who expects positive improvements in
compatibility in the long run.

\subsubsection{Confusion About Plans}

It seems that several updates to the plan have caused
uncertainties among the developers.
For example, R33 lost trust in the deadlines and paused the migration. 
Only 5 projects are satisfied with the information
given by browser vendors whereas 8 projects are not
satisfied. One of these unsatisfied projects does not understand the Firefox plan.
Additional 4 projects do not
understand the Chrome plan.

Our participants are also unsure about the motives for the changes introduced by
Mv3. Some, like R09 and R19, add that they do not understand why some
APIs must be removed. On the contrary, R14 approves access to the content of web
requests as it gives very high powers to webextension developers. However,
Chrome does not remove access to the content of the web requests, as that is
also possible by not blocking web requests (also stressed by R27). Even so, R14
is correct in the sense that malicious actors will have limited powers as they
would not be able to change the content of web requests and replies.

R19 speculates on motives behind the scene. According to the developer of R19,
the change to Mv3 benefits Google's business model.

\subsection{RQ3: Migration Process}
\label{sec:migration}

RQ3 deals with the migration period --- is it smooth or does it take a lot of
time?

Most projects that finished the migration by the time they responded
reported low impact on the code and consequently
invested only limited
development time to the migration: 3~projects invested
just days (up to a work-week), and one project needed a work month.
In contrast, R29 is an exception that reimplemented
well in advance despite
huge costs and months of working time.

Similarly,
projects estimating long migration time defer the migration.

\begin{itemize}

  \item Two projects expected short migration process but 
    had not invested any time to the
    migration before they answered our questionnaire. 
  \item Two projects had already invested several days and are expected to invest
    several more to complete the transition.
  \item One project had already invested less than one month and expected one
    more month to finish.
  \item Six projects expected multiple months for the complete transition (some
    did not disclose their expectations, others mentioned four months up to one
    year). These projects were in different phases of the transition, often,
    around the middle.

\end{itemize}

Hence, while we see projects with short and long expected migration times in both
groups, the projects with long expected migration times are more likely to be
inside the transition period.

We tracked the published version in the Chrome Web Store around the
migration deadlines (see Table \ref{tab:migration_table_respondents}).
At the beginning of June 2024, Google removed Featured badges from Mv2
webextensions. A Featured badge is manually given to projects that follow Google's best
technical practices and meet a high standard of user experience and
design~\citep{chrome_web_store_badges}.

\begin{table*}[h]
  \caption{Status of the participating projects around the Chrome deadlines to make the transitions}
  \label{tab:migration_table_respondents}
  \centering
  \begin{tabular}{|l|l|r|c|r|c|r|c|r|c|}
\multirow{5}{*}{\rotatebox{90}{\parbox{1.5cm}{Participant project ID}}} & \multirow{5}{*}{\rotatebox{90}{\parbox{1.5cm}{Expected migration duration}}} & \multicolumn{2}{c|}{Before badge} & \multicolumn{2}{c|}{After badge} & \multicolumn{2}{c|}{Chrome 127} & \multicolumn{2}{c|}{Chrome 127}\\
 &  & \multicolumn{2}{c|}{deadline} & \multicolumn{2}{c|}{deadline} & \multicolumn{2}{c|}{early release} & \multicolumn{2}{c|}{stable release}\\
 &  & \multicolumn{2}{c|}{2024-05-24} & \multicolumn{2}{c|}{2024-06-04} & \multicolumn{2}{c|}{2024-07-17} & \multicolumn{2}{c|}{2024-07-23}\\
\cline{3-10}
 &  & \multicolumn{1}{c|}{Mv?} & \multicolumn{1}{c|}{Featured} & \multicolumn{1}{c|}{Mv?} & \multicolumn{1}{c|}{Featured} & \multicolumn{1}{c|}{Mv?} & \multicolumn{1}{c|}{Featured} & \multicolumn{1}{c|}{Mv?} & \multicolumn{1}{c|}{Featured}\\
 &  & \multicolumn{1}{c|}{} & \multicolumn{1}{c|}{badge} & \multicolumn{1}{c|}{} & \multicolumn{1}{c|}{badge} & \multicolumn{1}{c|}{} & \multicolumn{1}{c|}{badge} & \multicolumn{1}{c|}{} & \multicolumn{1}{c|}{badge}\\
\hline
R04 & Not affected & 3 & YES & 3 & YES & 3 & YES & 3 & YES\\
R13 & Several hours & 3 & NO & 3 & NO & 3 & NO & 3 & NO\\
R23 & Days & 3 & YES & 3 & YES & 3 & YES & 3 & YES\\
R26 & Days & 3 & YES & 3 & YES & 3 & YES & 3 & YES\\
R27 & Days & 2 & YES & 2 & NO & 2 & NO & 2 & NO\\
R33 & Days & 2 & YES & 2 & NO & 2 & NO & 2 & NO\\
R07 & 1--2 months & 3 & YES & 3 & YES & 3 & YES & 3 & YES\\
R03 & Months & 2 & NO & 3 & NO & 3 & NO & 3 & NO\\
R10 & Months & 2 & YES & 2 & NO & 2 & NO & 2 & NO\\
R29 & Months & 3 & YES & 3 & YES & 3 & YES & 3 & YES\\
R01 & 12 months & 2 & YES & 3 & YES & 3 & YES & 3 & YES\\
R34 & Very long & 3 & YES & 3 & YES & 3 & YES & 3 & YES\\
R18 & Undisclosed & 3 & YES & 3 & YES & 3 & YES & 3 & YES\\
R31 & Undisclosed & 3 & YES & 3 & YES & 3 & YES & 3 & YES\\
R06 & Not planned & 2 & YES & 2 & NO & 2 & NO & 2 & NO\\
R14 & Not planned & 2 & NO & 2 & NO & 2 & NO & 2 & NO\\
R16 & Not planned & 2 & YES & 2 & NO & 2 & NO & 2 & NO\\
R17 & Not planned & 2 & NO & 2 & NO & 2 & NO & 2 & NO\\
R19 & Not planned & 2 & YES & 2 & NO & 2 & NO & 2 & NO\\
\end{tabular}
\end{table*}

Almost all projects that advertised that they were going to migrate managed to
publish a Mv3-compatible version of their extension. R10 expressed that
he got busy and could not finish the migration in time. R27 and R33 expected a
smooth transition of just a few days, yet they did not update. A probable reason
is that R27 waits until Firefox supports Mv3 properly, as expressed in
the answer. R33 explained that they lost faith in the deadlines. Consequently,
more than ten thousand Chrome Web Store users lost their webextensions.
R01 migrated a short time before the announced Featured badge deadline and
managed to keep the badge.

Let us focus on internals that might complicate the migration and how
the participants deal with the
challenges arising from the change of the APIs.

\subsubsection{Web Request API}

14 participating webextensions use Web Request API. However,
3 of them use the API just for non-blocking
purposes, so 11 participating extensions need blocking
Web Request API.
These participants approached the challenge of migration differently.

Two proxy managers, one page content sanitizer, and one tracker blocker claim that the
functionality can be reimplemented using Mv3 APIs without functionality
loss. 
However, one of the proxy managers explained that the implementation suffers
from bugs.
Additionally, the project maintainer of the tracker blocker explained that many
blockers want to provide feedback to users on what they did to each page. As the
declarative net request API does not provide feedback, they keep using web
request API in parallel to the declarative net request API and estimate what
actions should have happened through the declarative net request APIs.

One blocker (R31) removed a feature depending on blocking web request API and,
consequently, lost functionality during the migration.

An adblocker, an authentication tool, and two cookie managers state that the migration is
impossible. One project did not explain why it uses the blocking web request API, other
3 participating
webextensions need to modify the requests. The projects decided that it is
better not to offer the extension than to offer an extension that does not fully
work.

One project did not reveal its approach to the problem, and one project was
still in the process of deciding how to approach the issue.

It looks like the declarative net requests API is suitable for some types of
webextensions (proxy managers, page content sanitizers, some blockers) but
other types of
webextensions cannot use the API to reimplement the original behavior (mainly
various blockers, cookie managers, and authentication tools).

\subsubsection{Storing state}

Another common problem is the state stored in background
scripts. While Mv2 webextensions can use background variables to store
states, Mv3 webextensions need to access asynchronous API to access
storages offered by browsers. Consequently, other scripts might run while the
browser performs the asynchronous calls. The majority of participating
webextensions (18\unskip) stores state in background
scripts. 8\unskip~projects answered that they do not store
state.

Most projects find handling state challenging or cannot make it work
(see Table~\ref{tab:15b}). Two projects keep
confidential information that should stay in RAM in the browser storage. R28 underlines the
issue of storing data on private tabs. However, the content
of variables in background scripts of Mv2 webextensions could have been
moved to swap as well. So here, the better solution should be to provide an API
that can safely store confidential content. Two projects expressed that the
storage is slow. One project depends on an external library, so it needs to
wait until the library is migrated or rewrite the code. One projects worries
that messages sent from other parts of the extension 
do not start the worker.

\begin{table}[h]
  \caption{Is migration to background workers smooth?}
  \label{tab:15b}
  \centering
  \begin{tabular}{lr}
Response & Webext. count\\
    \hline
Yes & 1\\
General challenge & 7\\
New solution does not work well & 6\\
Undisclosed & 6\\
\end{tabular}
\end{table}

Only 1 project feels impacted by the asynchronous calls of
the storage access. 2 more projects expressed
a small impact.

\subsection{Comparison to All Selected Projects in Chrome Web Store}
\label{sec:comparison_to_cws}

\begin{figure}[h]
  \centering
  \includegraphics[width=\linewidth]{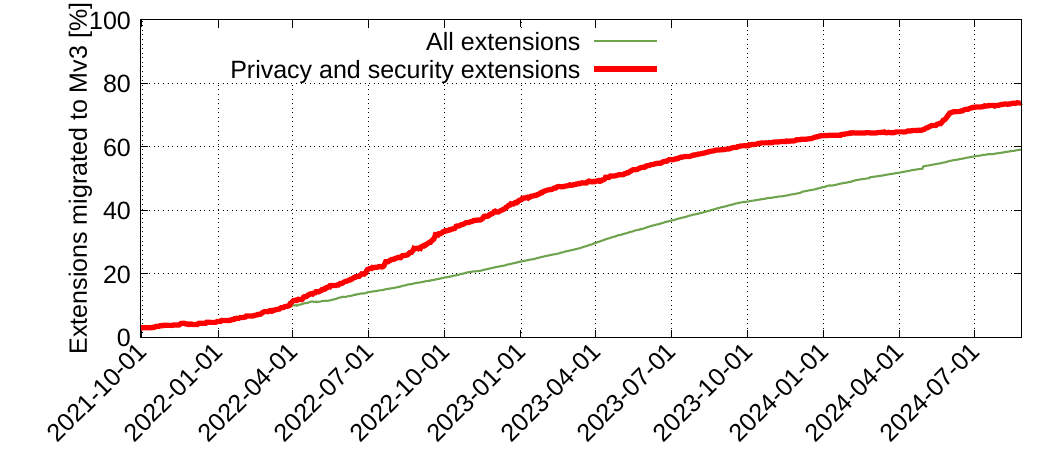}
  \caption{Share of webextensions that migrated to Mv3.}
  \label{fig:manifest_transition}
\end{figure}

Recall that  we found 1,110 webextensions of interest in Chrome
Web Store
(see Section~\ref{sec:searching_participants}).
Table~\ref{tab:chrome_web_store_all_candidates} summarizes the status
of the extensions at 2024-08-26, and Fig.~\ref{fig:manifest_transition} shows the
progress of migration in time with the comparison to all extensions.

\begin{table}
  \caption{Migration status of all projects we selected during the search for
  participants (2024-08-26).}
  \label{tab:chrome_web_store_all_candidates}
  \centering
  \begin{tabular}{lrr}
    Version in store & Count & Percentage \\
    \hline
    Mv2 & 266 & 26.0\,\% \\
    Mv3 & 756 & 74.0\,\% \\
    Removed from store with Mv2 & 19 & 21.6\,\% \\
    Removed from store with Mv3 & 69 & 78.4\,\% \\
\end{tabular}
\end{table}

\begin{figure}[h]
  \centering
  \includegraphics[width=\linewidth]{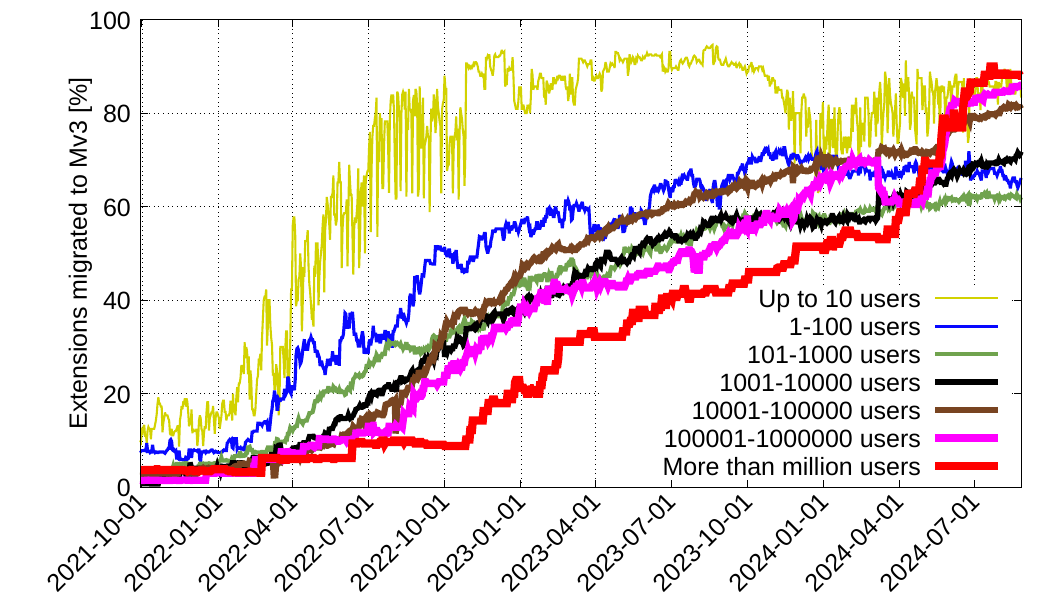}
  \caption{Share of privacy and security webextensions (with respect to their user count) that migrated to Mv3.}
  \label{fig:transition_bins}
\end{figure}

Figure~\ref{fig:transition_bins} shows the migration status of the selected
privacy and security webextensions in time; the webextensions are grouped by the
user count\footnote{See footnote~\ref{ftn:users_count} and note that
users with deactivated webextensions, for example, because their
browser no longer supports Mv2 webextensions, are counted by Chrome
Web Store while their browser checks for updates.}.
At the beginning of the timeframe, small webextensions were more
likely to employ Mv3. This trend is likely caused by new projects
starting directly in Mv3 and avoiding transition. Projects with bigger
user count were more likely to wait. This trend is especially visible for
webextension with more than a million users. The transition rate increased
significantly in about three months before the transition deadline. 44 out of 50
projects with more than a  million users migrated by the end of August 2024.
Nevertheless, six projects with a huge user count did not migrate in time.

Our study represents a
much higher share of projects with troubles during migration (see
Tab.~\ref{tab:migration_table_respondents}) compared to the
overall picture. Even so, we consider alarming
that about one quarter of
projects in our research group did not migrate.
The analysis of responses to our questionnaire provides explanation why
projects decided not to migrate.

\subsection{RQ4: Benefits of Mv3}

The last research question 
deals with possible benefits of Mv3 to webextensions.

\subsubsection{Is Mv3 Safer Than Mv2?}

Some of the projects in our group of likely privacy- or security-related
extensions were removed by July 2023 (see
Tab.~\ref{tab:chrome_web_store_all_candidates} and
\ref{tab:chrome_web_store_removed_ext}). The removal might have been initiated
by the owner of the webextension who decided to stop offering the code. Often,
the removal is caused due to fraudulent purposes of the webextension. As a
significant portion of the removed webextension for fraudulent behavior were
published with Mv3, we question if the goal of designing Mv3
as a safer replacement holds in practice.

\begin{table}
  \caption{Removal reason as listed by Chrome-stats.com.}
  \label{tab:chrome_web_store_removed_ext}
  \centering
  \begin{tabular}{lrr}
    Reason & Mv2 & Mv3 \\
    \hline
    Bundled with unwanted software & 0 & 2 \\
    Policy violations & 3 & 7 \\
    Malware & 2 & 19 \\
    Unknown & 14 & 41 \\
\end{tabular}
\end{table}

\subsubsection{Lack of benefits to users}

A majority of our participants do not see a benefit for the users in the
migration to Mv3 (see Table~\ref{tab:16}).

\begin{table}[h]
  \caption{Will Manifest V3 make your extensions better for your users?}
  \label{tab:16}
  \centering
  \begin{tabular}{lr}
Response & Webext. count\\
    \hline
Yes & 2\\
Yes (only in Firefox for Android) & 1\\
No & 15\\
Uses Mv3 from start & 1\\
\end{tabular}
\end{table}

Both participants who think that Mv3 will improve their extension for users see
the main benefit in making their webextension less performance-heavy.  For example, R07 finds the
new declarative net request API less performance-heavy. R16 considers Mv3
to have made webextensions more accessible to users. R23 highlights the
benefit of Mv3 in preserving the state of the extension among restarts
of the browser caused by the operating system.

Some of the projects that do not think that Mv3 will make their
webextension better to their users fear future changes. Two projects are aware
that their Mv3 webextension is buggy. Some recall that some removed APIs (like
blocking web requests API) do not have a full replacement, meaning some extensions
cannot be fully migrated. R26 needed to remove a feature as he did not
understand how to implement the functionality using Mv3 APIs.

R01 expressed: \enquote{The gist of Mv3 work is that it is work that is almost entirely 
invisible to the user if we do it right. The users will only notice that 
something changed if something no longer works right or is missing. 
There are almost no benefits to users, and certainly none the vast 
majority will ever notice. When we work on MV3, whether for Chrome or 
Firefox, we are not working on making our webextension work better for its 
users. We are running to stay in place.} R29 adds that migration to Mv3
is the biggest project they need to undertake, yet, the migration is
completely opaque to users.

\subsubsection{Content scripts}
\label{sec:content_scripts}

\emph{Content scripts} run in the context of a visited page and have access
to its document object model. Consequently, they can manipulate the content of
the page, including removing, replacing, or extending some parts of the
visited page or the page as a whole.

Some webextensions need to inject content scripts that must run before page scripts start
running. For example, such content scripts can change the implementation of some
APIs to provide modified or fabricated values~\citep{js0,jshelter}. Browsers
offer \verb|scripting.RegisteredContentScript| API that allows injecting content
scripts to the
page\footnote{\url{https://developer.mozilla.org/en-US/docs/Mozilla/Add-ons/WebExtensions/API/scripting/RegisteredContentScript}}.
However, this API is not available for Mv2 webextensions in
Chrome\footnote{\url{https://developer.mozilla.org/en-US/docs/Mozilla/Add-ons/WebExtensions/API/scripting}}.
Nevertheless, this API is not the only way to inject content scripts.
For example, some projects like NoScript Commons Library solve the issue of
reliable injection for Mv2 webextensions using other APIs~\citep{jshelter}.

Hence, we asked the participants if they need to run content scripts in advance
and if they are confident that their scripts indeed run in time in Mv2
and Mv3. Table \ref{tab:12interest} shows that only six participants
replied, most were not sure that they inject scripts in time in their Mv2
extension and no one is sure in the Mv3 version. This raises serious
questions about the reliability of the extensions. While the developers are not
sure that the webextensions work as they should, should users depend on such
webextensions? Do users understand the risks? We leave answering these questions
to future research.

\begin{table}[h]
  \caption{Are developers confident that content scripts run on time? (\emph{-} means undisclosed, \emph{?} means unsure.)}
  \label{tab:12interest}
  \centering
  \begin{tabular}{llll}
PPID & Mv2 & Mv3 & Purpose of content scripts\\
    \hline
R01 & - & - & Replace some DOM parts \\ & & & and modify APIs available \\  & & & to page scripts\\
R15 & No & ? & Modify APIs available to \\  & & & page scripts\\
R25 & - & - & Replace some DOM parts\\
R28 & Yes & No & Store settings\\
R31 & No & No & Replace some DOM parts\\
R35 & Yes & No & Replace some DOM parts\\
\end{tabular}
\end{table}

The developer of R16 and R17 further adds that as he is not confident that
content scripts run in time, he gives up all efforts on extensions that need
such a functionality.  The lack of suitable APIs likely prevents some
extensions from being developed, as potential developers are not persuaded
that the extensions would be reliable.

A related issue is that Firefox suffers from a long-standing
bug 1267027\footnote{\url{https://bugzilla.mozilla.org/show_bug.cgi?id=1267027},
the bug was
opened 2016-04-23 and is not fixed as of 2024-07-17.} that prevents content
scripts from injecting scripts by inserting \verb|script| elements to the page in the
presence of Content Security Policy that disallows such scripts. This means that
the author of the visited page can intentionally or unintentionally prevent
webextensions from performing their intended functionality if the author of the
webextension is unaware of the bug. For example, GitHub contains tens of
issues referring to the Firefox
bug\footnote{\url{https://github.com/search?q=https\%3A\%2F\%2Fbugzilla.mozilla.org\%2Fshow_bug.cgi\%3Fid\%3D1267027&type=issues}}.

Table \ref{tab:12+13} shows the replies of our respondents who support Firefox.
According to the responses of the participants, the majority is not affected.
However, only one participant disclosed the mitigations employed by the
webextension --- the webextension modifies the Content Security Policy. As the
policy is primarily intended to prevent bugs like cross-site scripting, lowering
the policy might expose users to unintentional security risks~\citep{helping_hindering}.

\begin{table}[h]
  \caption{Webextensions affected by Firefox bug 1267027}
  \label{tab:12+13}
  \centering
  \begin{tabular}{lr}
Response & Webext. count\\
    \hline
Affected & 2\\
Undisclosed & 3\\
Do not know & 1\\
Not affected & 9\\
\end{tabular}
\end{table}

We see two potential approaches to solve the problem. Firstly, Firefox should
fix bug 1267027, so that content scripts work predictably and compatibly
with Chromium-based browsers. Alternatively, browsers should provide
configurable and reliable script injection mechanism without
side-effects\footnote{See, for example,
\url{https://github.com/w3c/webextensions/issues/103} for a related issue at
W3C.}. The latter approach is better as it does not leave
artifacts that can be misused for
browser fingerprinting~\citep{xhound,extend_not_extend}.

\subsubsection{APIs Missing in General}
\label{sec:missing_apis}

Finally, we asked participants about the APIs they think are missing in general.
Browser vendors can be inspired by these APIs
to improve the browsers extensibility
and enable developers to create webextensions that are not possible today.

Some developers highlighted browser compatibility. For example, they mentioned
that some APIs behave differently in Firefox and Chromium-based browsers or that
Firefox has additional APIs like the DNS
API\footnote{\url{https://developer.mozilla.org/en-US/docs/Mozilla/Add-ons/WebExtensions/API/dns}}
or event pages\footnote{W3C issue tracker on the topic: \url{https://github.com/w3c/webextensions/issues/134}}.

Other developers suggested improvements to existing APIs. For example, the web
request API provides the browser tab ID, but it does not provide the URL of that
tab. Currently, webextensions need to develop workarounds to obtain the URL. The
extended API could simplify the code of webextensions. Another example is to
provide a dedicated API to change web request headers. Some provided further
examples of improvements to web requests, such as allowing blocking based on
resolved IP addresses and information from TLS certificates. Others mentioned
improvements to the storages browsers offer, such as an API that removes records
with specific keys.

Participants also proposed new APIs:

\begin{itemize}
  \item to modify built-in APIs,
  \item to insert DOM overlay content not accessible to page scripts to prevent
    webextension fingerprintability~\citep{xhound,extend_not_extend},
  \item to search all DOM, including all shadow DOMs for certain elements, for
    example to detect page artifacts of interest hidden in shadow DOMs,
  \item to improve consistent processing of the top level documents and
    documents displayed in iframes,
  \item to enable secret handling,
  \item high-performance storage API,
  \item to improve icons displayed in dark themes,
  \item to allow background and content scripts to open the webextension popup
    window.

\end{itemize}

\subsection{Other lessons learned}
\label{sec:insights}

Some developers decided to take another path instead of migrating their
webextension to Mv3.
R06 developed 6 extensions and is bothered by several migrations that browser extensions experienced in
the past. 
R06 had hoped that webextensions would unify the code base of browser
extensions which was originally successful for R06. However, R06 feels that
such an opportunity was lost with Mv3 as it requires learning from
fragmented information and introduces another platform to migrate to.
As a result, R06 reimplemented
the most important
functionality of the maintained extensions as a user script of
Greasemonkey\footnote{\url{https://www.greasespot.net/}}, a webextension that
allows customization of the way webpages look and allows sharing user
scripts\footnote{\url{https://wiki.greasespot.net/User_Script_Hosting}}. Effectively, R06 added
another layer that removes the direct dependency on webextension APIs
as long as Greasemonkey continues to work. The cost is that some
functionality is not possible to implement in Greasemonkey.
Additionally, as the move to Greasemonkey user scripts is not expected by
browser vendors, each user needs to manually replace the original webextension with
Greasomonkey user script.


In one case, we reached an academic researcher who merely stated that the
extension is not meant to be used by the general public and should be removed
from the public store. This raises questions of how many similar projects are
available in the stores, if installing such abandoned projects would introduce
any danger to the user, and how users can be protected from installing
potentially harmful extensions that are not developed anymore.

\section{\uppercase{Conclusion}}
\label{sec:conclusion}

Mv3 was supposed to promote secure and privacy-respective
webextensions~\citep{google_mv3_2018} without limiting their powers. While we
have seen some positive effects of the migration, like one project adding the
support for a new browser and another project unified code bases,
most participants reported the need to redesign
their extensions without a positive impact on the users, dropped support for
some browsers, or need to maintain separate code bases for the supported
browsers.

The majority of participants reported a lack of benefits to users; only a few of
the projects considered Mv3 to improve the performance of their
webextensions. Others question this benefit. For example, one of our
participants consumes additional resources to keep the background scripts
running. Hence, such extensions consume more electrical power (possibly from a
battery) compared to the Mv2 version.

Our research evaluates the reasons why webextension developers are hesitant to
change. Important APIs are missing or are not mature and stable enough.
Additionally, Mv3 lacks new features and possibilities and does not
motivate developers to switch to the new APIs. Cross-platform compatibility is
another observed problem. Projects switch to separate code bases for
Chromium-based browsers or completely give up support for these browsers.

Although some projects report short migration time, others observe problems,
including uncertainties in the guarantees that content scripts run in time, lost
functionality due to missing APIs, slow and unsafe means to store state, and
bugs. Some types of webextensions like proxy managers and page content
sanitizers are affected lower compared to various blockers, cookie managers, and
authentication tools. Participants also worried about the implications of
storing confidential information (e.g. cryptographic secrets, information from
private browser tabs) in the webextension storages.

Although subsection~\ref{sec:migration_problems} shows that one reason that
prevents developers from migration was a lack of benefits, R15 raised an
important argument. Given the webextension can be published by malicious actors,
limiting their powers can be considered as a good future path.

\section*{\uppercase{Acknowledgements}}

This project was funded through the NGI0 Entrust Fund, a fund established by NLnet with financial support from the European Commission's Next Generation Internet program, under the aegis of DG Communications Networks, Content and Technology under grant agreement No 101069594 as JShelter Manifest V3 project.
This work was partly supported by the Brno University of
Technology grant FIT-S-23-8209.

\bibliographystyle{apalike}
{\small
\bibliography{literature}}

\appendix

\section{Questionnaire}
\label{appx:questionnaire}

We send an e-mail to each potential participant with the following questions:

\begin{quote}
(1) What is the name of your project?

(2) Do you provide answers for yourself or for the project as a whole?

(3) How many users does your project have? Rough numbers are sufficient. What kind of users benefit
from your extension most? Please explain how they benefit.

(4) Is your project affected by the migration to Manifest V3? Please, explain how. For example,
explain features that you plan to add to your extension due to Manifest V3 and explain features
that will not be possible to keep or that will need to be rewritten. If you have a plan, please,
provide us its milestones and current status.

(5) Are the APIs offered by the Manifest V3 stable and mature enough? Please explain your view.

(6) Is the documentation on the Manifest V2 and V3 of sufficient quality? Do you miss any tool
for debugging or experimenting with the APIs? Please, explain.

(7) What browsers does your extension support? Did you maintain separate versions for specific
browsers before the migration to Manifest V3 and during the migration? Do you plan to have separate
versions after the migration or if the migration already finished, do you have different versions for
specific browsers? Please explain.

(8) Do you think that Manifest V3 improves the compatibility of webextensions across browser
vendors? Please explain.

(9) Are you content with the information given by browsers vendors? Is it clear what, how, when, and
why needs to change? Please explain.

(10) Please, provide estimates of working months that you or your project already invested into the
migration.

(11) Please, provide estimates of working months that you or your project will need to invest into
the migration.

(12a) Does your project use content scripts? Please explain how and why?

(12b) If your project deploys content scripts, does the extension need to execute them before page
scripts start running? Please explain.

(12c) If your project deploys content scripts that need to run before page scripts start running,
are you confident that they actually run in time? Please explain you view regarding the APIs
available in Manifest V2 and V3 separately.

(13) Is your extension in any way affected by
\url{https://bugzilla.mozilla.org/show_bug.cgi?id=1267027}?
Please explain how and your approach to minimize the impact on the users.

(14) Does your extension use WebRequest API? Please explain why and how. Is that functionality
affected by the migration to Manifest V3? Please explain how you tackle the problem.

(15) Have you used Manifest V2 background scripts to store state? If so, is the migration to workers
smooth? Have you encountered any pitfalls? Please explain. Does Manifest V3 force you to
asynchronous calls that might affect the functionality of the extension? Please explain.

(16) Do you think that Manifest V3 will make your extensions better for your users? Please explain your
view.

(17) Are you aware of the existence of W3C's Web Extensions Community Group (WECG)? Do you have any
experience with the group? Please, let us know.

(18) Is there any API that you think is missing for web extensions? Please explain the needed
functionality and purpose of the API.

(19) Do you have anything important that you think we missed in the questionnaire? Please, let us know.

(20) Do you want to be contacted back if we have further questions?

(21) Do you want to be notified about the paper?
\end{quote}

Note that we have not evaluated question 17 as we later asked participants of
WECG to provide answers in the hope to learn benefits of Manifest v3.

\end{document}